# Flexible focal plane arrays for UVOIR wide field instrumentation

Emmanuel Hugot [a], Wilfried Jahn [a], Bertrand Chambion [b], Gaid Moulin [b], Liubov Nikitushkina [b], Christophe Gaschet [a, b], David Henry [b], Stéphane Getin [b], Marc Ferrari [a], Yann Gaeremynck [a, b]

[a] Aix Marseille Université, CNRS, LAM (Laboratoire d'Astrophysique de Marseille) UMR 7326, Marseille, France
[b] Univ. Grenoble Alpes, CEA, LETI, MINATEC campus, F38054 Grenoble, France.

**ABSTRACT**

LAM and CEA-LETI are developing the technology of deformable detectors, for UV, VIS or NIR applications. Such breakthrough devices will be a revolution for future wide field imagers and spectrographs, firstly by improving the image quality with better off-axis sharpness, resolution, brightness while scaling down the optical system, secondly by overcoming the manufacturing issues identified so far and by offering a flexibility and versatility in optical design. The technology of curved detectors can benefit of the developments of active and deformable structures, to provide a flexibility and a fine tuning of the detectors curvature by thinning down the substrate without modifying the fabrication process of the active pixels. We present studies done so far on optical design improvements, the technological demonstrators we developed and their performances as well as the future five-years roadmap for these developments.

## 1. THE SMALL WORLD OF CURVED DETECTORS AND ITS HIGH POTENTIAL IMPACT ON INSTRUMENTATION

The emergence of curved detectors, first proposed by Ko et al in their Nature paper [1], certainly represents the major disruptive technology for imaging systems that will come up in a near future. But why are curved detectors so full of potential and not yet off-the-shelf components? Two main reasons: affordable manufacturing process must be developed, as well as the market must be developed in parallel. The first commercial products have been announced recently by Sony (Itonaga 2014 [2]).

Such a breakthrough device has a very high potential impact for astronomy. By directly correcting the field curvature in the focal plane of wide field instruments and telescopes, the use of curved detectors allows a drastic reduction of the optical systems complexity and will increase the transmitted flux and then reduce the exposure time by a significant factor. Also, less optics means less misalignments and instrumental errors, obvious increase of the stability of instruments and PSF homogeneity (key point for applications such as astrometry or galaxy morphology study), less calibration time, less dependence to the environmental condition: the simpler the merrier.

At the international level, several developments of curved detectors have been pursued by Sarnoff (Swain et al [2]), Stanford (Rim et al [4], Dinyari et al [5]), University of Arizona (Lesser [6]) and JPL (Nikzad et al [7]). Different techniques for the bending of CCD and CMOS detectors have been proposed and the prototypes give exquisite results, with a very low degradation of the detector performance in terms of dark current, shift charge or noise, and a filling factor above 80%.

### 1.1 Static curvature

As explained by Rim et al or Fendler & Dumas [8][9][10], the curvature of focal plane arrays leads to a drastic simplification of the optical systems by offering a new parameter in the optimizations. The field curvature, so called the Petzval surface, no more has to be compensated by a set of complex optics flattening the field.

One of the nicest example of the impact of curvature in the field is the KEPLER focal plane array, made of 21 flat large format detectors disposed on a highly curved plate (Figure 1). This solution of discrete curvature however presents some drawbacks for high angular resolution imaging: the optimization of the optical quality is done over zones, and the PSF shape is no more homogeneous over the local field.

R&D activities have been launched by ESO in the field of astronomy, in collaboration with the University of Arizona, for the development of large format VIS curved detectors with a consequent curvature (Iwert et al [11], Figure 1). In that case, the manufacturing process has been fully developed and delivered a perfectly working component suitable for astronomical applications.

In France, the activity undertaken at CEA-LETI for now ten years in this field has produced major results, the most impressive one being the development of near infrared curved detectors (HgCdTe detectors, also called CMT or MCT detectors), operational at 80K (Tekaya et al 2014 [12]) with a 100% filling factor. This achievement was done by adapting the curving process developed by Dumas and Fendler published in 2012 for bolometric arrays working at room temperature and at a wavelength of 11µm (Figure 1). This group also worked on the optical design optimizations of the proposed instrumentation for the E-ELT (Fendler+ [8]) by use of curved focal planes, demonstrating a global increase in performance with the use of less component, less complex. They also demonstrated that a major issue in their process was the curving of convex detectors, for mechanical reasons, that degrades the shape and the performance.

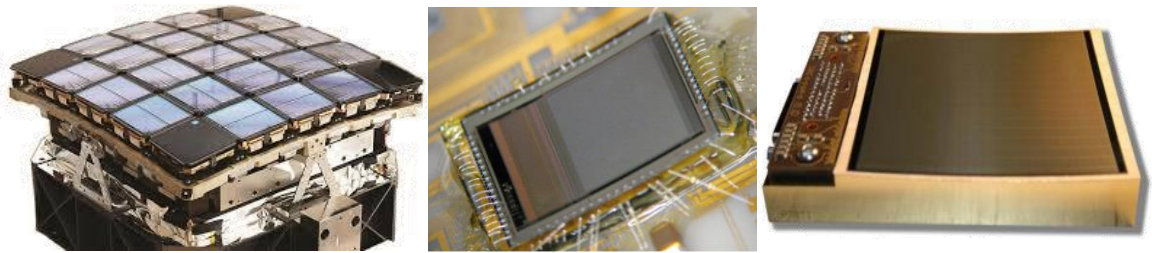

Figure 1. Left: The Kepler Focal plane made of flat detectors on a curved support. Middle: A curved bolometric array from CEA-LETI. Right: Curved array designed for astronomy.

**1.2 Dynamic curvature**

During the last four years, the French collaboration with LAM and CEA-LETI, allowed the realization of the first prototype of deformable VIS detector associated to a wide field zoom optical system. In that case, we consider active mirrors as substrates, accurately controllable in shape, on which the functional thin detectors are glued. While the precisions required for the accurate bending and control of active mirrors is below the micron, the precision required for detectors is way less demanding, only related to the depth of field.

The active mirrors zoo developed at LAM the last years is perfectly suitable for this application and offers a brand new component: deformable detectors, which shape can be controlled actively during operations and observations. This component has a very high potential in terms of valorisation, technology transfer and trans-disciplinary applications. Wide field optical systems can be proposed with much less components, resulting in lightweight cameras with a high impact on embedded imaging systems in several fields (defence, Earth imaging, biomedical optics, commercial cameras …).

We designed and optimized a demonstrator of very wide field zoom system (170°). Figure 2 shows the comparison between an existing patented design (Canon [13]) and the one we proposed. For an equivalent performance, the gain is obvious: 9 lenses instead of 14, only spherical surfaces, 3 materials instead of 10, equivalent performance. This outstanding result is only made possible by adapting the detector curvature along the zoom range.

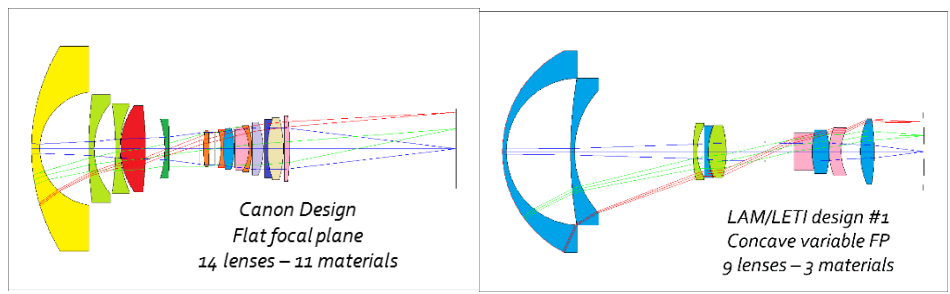

Figure 2. Comparison of two fish-eye objectives for commercial cameras (8-15mm at constant F/4). Left: existing patented design using a flat detector and a complex set of optics and materials. Right: our demonstrator using a variable curvature detector, with much less optics and three times less different materials, for an equivalent performance. Nice to have, this last system is also telecentric and uses only spherical surfaces, making it very tolerant to misalignments.

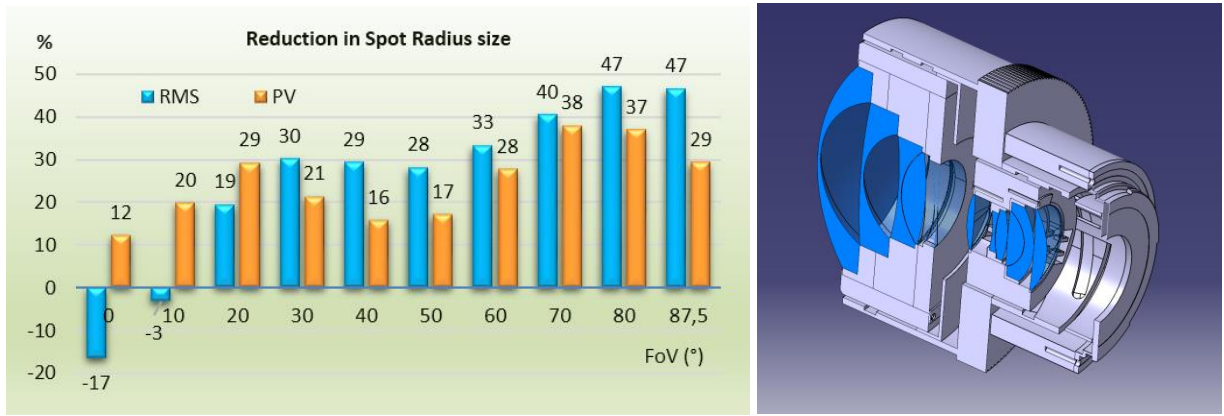

Figure 3. Left: Reduction of the spot radius, due to the use of the curved detector, showing very quick improvement up to 30% for a FoV of 60deg, and nearly 50% at the edge of the 175deg field of view. Right: CAD model of the prototype currently under integration at LAM.

The plot on Figure 3 (left) demonstrates the immediate gain on the optical quality we get by making use of a curved detector over a wide field. We note a quick improvement up to 30% in terms of spot size at 40 deg FoV, and nearly 50% improvement at 175deg.

This improvement on the optical quality comes along with a reduction of the surfaces counting, as well as a complete removal of any aspherical surface. This exquisite result motivated us to start the realization of a prototype for a full performance characterization.

For an efficient demonstration, we will likely use a high potential of 3D-printing for all the mechanical parts, as well as for the deformable substrate holding the sensor.

The next section focusses specifically on the variable curvature sensor packaging.

## 2. TUNABLE CURVING PACKAGING

The T-CFPA (Tunable curvature focal plane array) technology helps to simplify the optical design but enhances the complexity on the sensor packaging. Let's now analyze the consequences on curving shape, sensor mounting, die attach material, wire bonding, mechanical behavior, and electro-optical performances while curving.

**2.1 Flexible package for CMOS image Sensor**

A standard CMOS image sensor structure consists in a silicon die, glued with a structural glue on a ceramic package. Electrical connections are wire bonded from the die to the ceramic surface and then, to the interconnection board thanks to a reflow process. On top, a glass cover is placed to prevent from mechanical or environment solicitations.

For the variable curvature detector that we are prototyping, the CMOS sensor chips are 20 by 23 mm square which are diced and thinned down to roughly 100 μm thick with specific grinding equipment, in order to make the sensor mechanically flexible.

To give the CMOS sensor a spherical shape with a variable radius of curvature, we choose to hold it on a Variable Curvature Mirror technology (VCM technology) developed at Laboratory of Astrophysics of Marseille (LAM) for astrophysics applications [14][14]. VCM technology consists in a circular membrane with a variable thickness distribution. A force $F$ applied at the center of the structure and perpendicularly to the membrane surface results in a perfect spherical bending of the top surface of the membrane with a radius of curvature depending on the force $F$. The membrane working principle is presented through a cross section representation in figure 4.

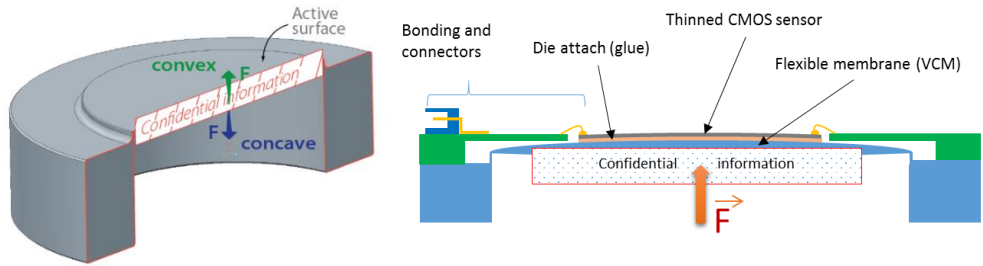

Figure 4. Cross section of the VCM structure, used as the flexible mechanical holder, and schematic representation of the T-CFPA technology packaging.

The T-CFPA packaging is detailed in figure 3. The thinned CMOS sensor is glued on the top surface of the VCM. Glue properties have been optimized in term of chemistry and thermomechanical properties to obtain as close as possible a spherical deformation when the force F is applied. We can apply either a pushing or pulling force, respectively for convex and concave CMOS curving. An interconnection board is placed close to the CMOS sensor to establish electrical connections from the die to the read-out circuit. The electrical connections between the sensor and the board consist in wire bonding. Due to the thickness of the sensor (~ 100µm), wire bonding process (including force, time, temperature and ultrasonic power) has been adjusted to prevent any damages or cracks at die level. A standard Zero Insertion Force (ZIF) connector (40 connections) is finally used to connect the read-out circuit.

## 2.2 FEA Model

We consider square plates of homogeneous materials permanently deformed into a spherical cap shape (Figure 4). Due to symmetry of the material, geometry and boundary conditions, only a quarter of the plate is modeled to reduce the model size.

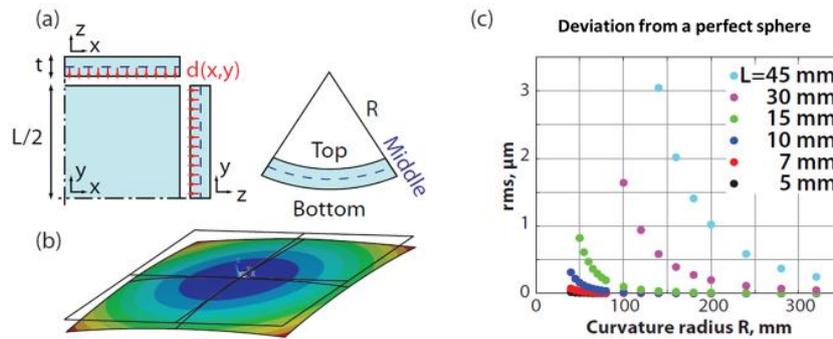

Figure 5. FEA model of a square plate. (a) One quarter of the plate, geometry and notations; (b) Spherically deformed middle surface of a plate. Vertical displacement field distribution Uz; (c) Deviation from a perfect sphere (RMS) as a function of imposed radius of curvature for a thickness t=20µm.

3D analyses are performed with the boundary conditions and are applied on nodes at (x,y,z)=(0,y,z) and (x,y,z)=(x,0,z). In addition, normal translations are constrained at the same nodes in order to prevent rigid-body motion during simulation and to avoid an underconstrained model. Furthermore, spherical deformation is generated by applying a vertical displacement d(x,y,R) to the nodes at the middle surface of the plate. Figure 4.b shows the distribution of vertical displacement Uz of the plate with size L=5mm, thickness t=20µm and radius of curvature R=60mm. To validate our numerical approach and to estimate the accuracy of our numerical results, we extract the obtained shape of the plate middle surface after deformation and its deviation from a perfect spherical shape. In Figure 4.c, the rms error to a perfect sphere is given as a function of imposed radius of curvature R for different plate sizes L. The figure shows that the error is about 1µm only. In the context of the current study, such small error is considered as acceptable. The reason the obtained shape is not perfectly spherical is that the plate is not infinitely stretchable.

**2.3 FEA results on silicon substrate flexibility**

Using FEA model, a parametric study is performed to investigate the influence of some geometric parameters on the overall capability of silicon to bend: plate size L, thickness of the plate t and radius of curvature R. We use following ranges of values: L ranges from 3 to 45mm, t – from 10 to 400µm and R – from 40 to 500mm. For each configuration, we focused on S1=S2 tensile stress at the top and the S3 compressive stress at the bottom (Figure 6) to extract the minimum allowable radius of curvature (safe region limit). For illustration, the stress distribution in the silicon plate (top and bottom) for R=200mm is presented in Figure 12. According to the stress fields, the top plane appears to be under equi-biaxial tension, and the bottom, under equi-biaxial-compression. Note that concerning the top plane stress field, a uniaxial compression field appears at the center of the edges. It could be the origin of a buckling effect as shown in a previous study on cooled IR sensor and due to a load-deflection response for thin-walled structures.

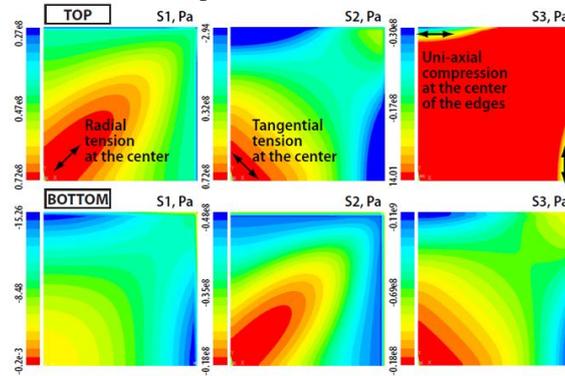

Figure 6. Principal stress distributions in the silicon plate of L=10mm and t=50mm subjected to spherical deformation of R=200mm. The distributions are plotted for the top and bottom plate faces.

Now, based on the plate mechanical behavior previously detailed, we plot on figure 7 a reference graph which consists in the minimum allowable curvature radius $R_{min}$ as a function of the silicon die thickness t, for several die sizes L. We can observe, as on Figure 6, that the bending limit is located at the top surface (equi-biaxial tension) of the plate for a convex shaping. As an example, referred to the black cross on figure 14, a 30x30mm silicon die must be thinned down to 160µm at least to be bent into a sphere of radius 150mm.

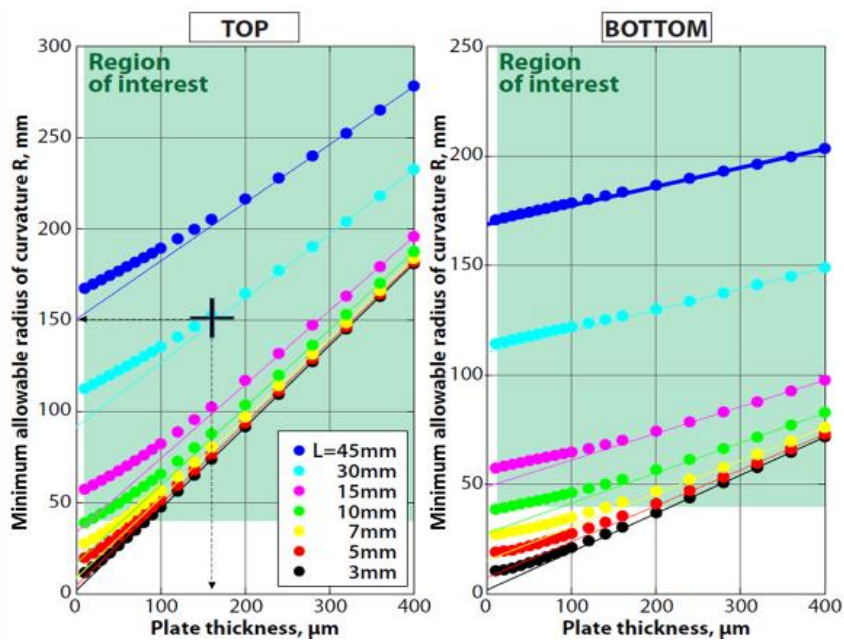

Figure 7. Results of the parametric study conducted for (100)-oriented silicon plate.

**2.4 FEA results on CMOS sensors**

This section introduces the results of FEA simulations of spherical deformation of a silicon-based monolithic CMOS sensor modeled by a two layers' laminate assembly as shown in Figure 16.a. The goal is to examine the mechanical behavior of curved layers and identify curvature limits in case of convex and concave bending modes. In addition, the effect of the silicon substrate thickness on the bending of the sensor is studied. Numerical simulations are performed for the following configuration: two layers' structure consisting of a silicon substrate of variable thickness $t_{Si}$ and a CMOS layer with thickness of 4µm. Two layers are bonded rigidly, i.e. without sliding. The sensor is assigned a constant length and width of L=5mm. The HEM properties of the CMOS layer are listed in Table 2. The spherical deformation is achieved by applying various radius of curvature at the bottom surface of the Silicon substrate. An example of the calculated vertical displacement field is shown on Figure 16.b in the case of both concave and convex-type bending modes.

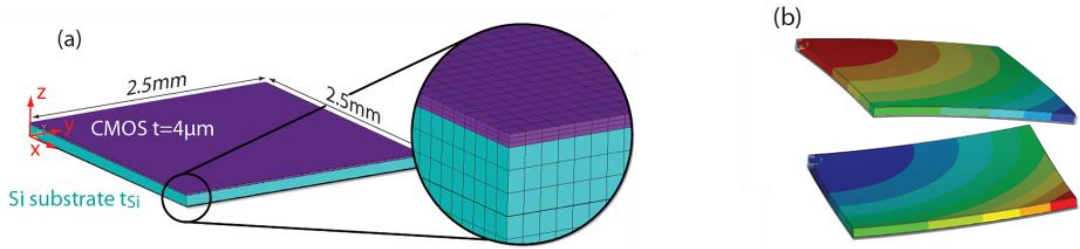

Figure 8. Modeling of a visible CMOS sensor. (a) Geometry and mesh definition; (b) An example of curved image sensor under convex-type deformation (upper image) and under concave-type deformation (bottom image).

We perform a parametric study to determine the effect of the silicon substrate thickness $t_{Si}$ on the minimum radius of curvature that the sensor can reach before failure of the active layer (upper surface of the CMOS layer). Specifically, the substrate thickness ranges from 10 to 210µm and the imposed radius of curvature varies from 40 to 400mm. For each configuration we check the maximum principal stress generated at the top surface of the sensor by comparing its magnitude to the flexural strength of the CMOS layer (see Table 2). Figure 8.a shows the minimum allowable radius of curvature as a function of the substrate thickness for both cases: concave and convex bending. The results indicate that a curved sensor has bigger capacity to bend under concave-type deformation than under convex-type deformation. However, by decreasing the substrate thickness below 50µm, small radius of curvature (close to R=30mm) can be achieved before the theoretical mechanical failure of the CMOS active layer.

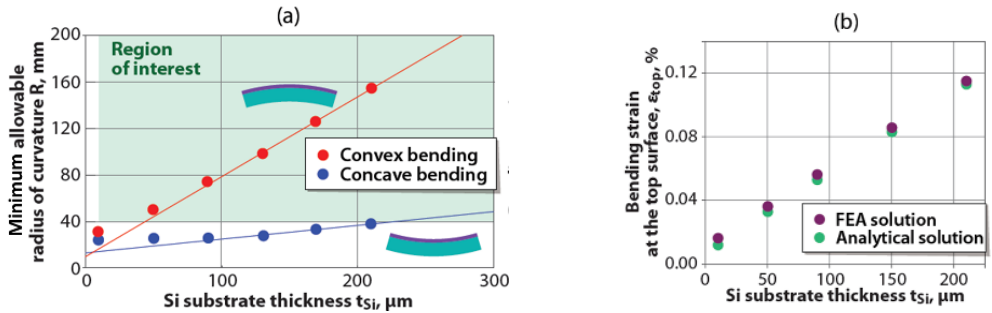

Figure 9. (a) Minimum allowable radius of curvature for a 5*5 mm sensor under concave- and convex-type deformations; (b) Bending strain at the top surface of the sensor as a function of the substrate thickness at fixed radius of curvature R=100mm.

The proposed numerical model is validated against the analytical solution derived in section IV.B for a bilayered plate under spherical deformation. Figure 17.b shows the magnitude of the bending strain calculated at the top surface of the convexly curved sensor as a function of the Si substrate thickness $t_{Si}$ for R=100mm. The simulation results are in good agreement with the theory. Thus, the results presented in Figure 17.a can be assumed as reliable. Nevertheless, further validation against experimental data is highly desirable with extensive experimental investigations of the bending characteristics of visible CMOS.

## 3. EXPERIMENTAL RESULTS

### 3.1 T-CFPA samples

In this section, we introduce T-CFPA samples on our flexible packaging approach (Figure 9). The packaging is based on a flexible membrane of 44 mm in diameter, printed on a prototyping tool. The thinned 20x23x0.1mm CMOS sensor is glued on the deformable substrate. Electrical connections are performed with a 25 µm diameter gold wire to validate the bonding process from a Printed Circuit Board (PCB) to the thinned CMOS sensor (figure 98 left). Due to the thickness of the sensor (~ 100µm), wire bonding process (including force, time, temperature and ultrasonic power) has been adjusted to prevent any damages or cracks at die level.

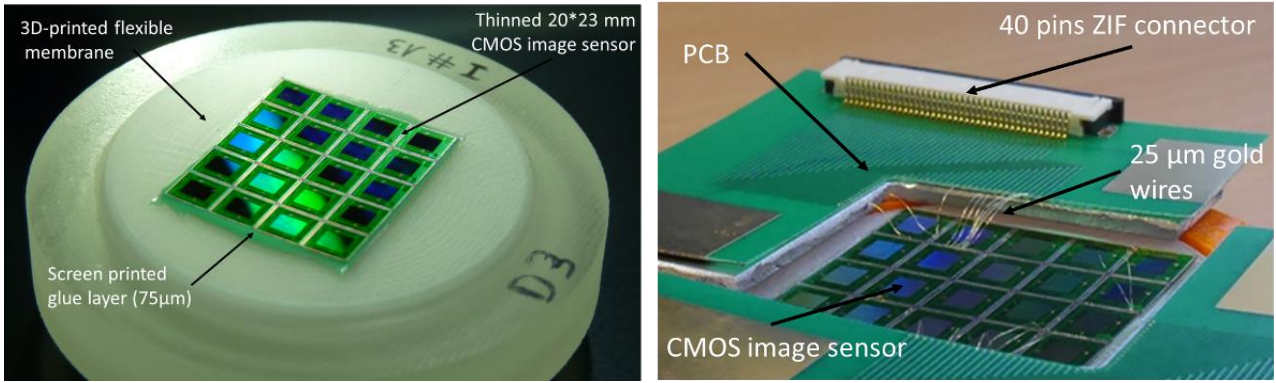

Figure 10. Left, Thinned Monolithic 20*23 mm CMOS image sensor glued on the 3D printed flexible membrane. Right: Thinned monolithic 20*23 mm CMOS image sensor in flat position, with its electrical wire connections to the external driving circuit

### 3.2 Bendable capabilities characterizations

The bending capabilities of the T-CFPA was first tested on bulk Silicon dies (20*23mm, 100µm thick). The top surface shape was characterized with an optical confocal profilometer. Surface shape characterizations are shown for flat position on figure 10 left, and for a curved position on figure 10 right. Curvature radius value is extracted via the bow value regarding the surface length. For flat position, the curvature radius should be +∞, but a 1380 mm radius is measured. This value could be due to a structure relaxation of the membrane material (polymer) after prototyping or due to the glue shrinkage while curing. For curved position, a down force is applied at the center of the membrane for concave deformation. It results in a 223 mm radius of curvature. Ten successive bending/unbending were performed to first estimate the reversibility of the system.

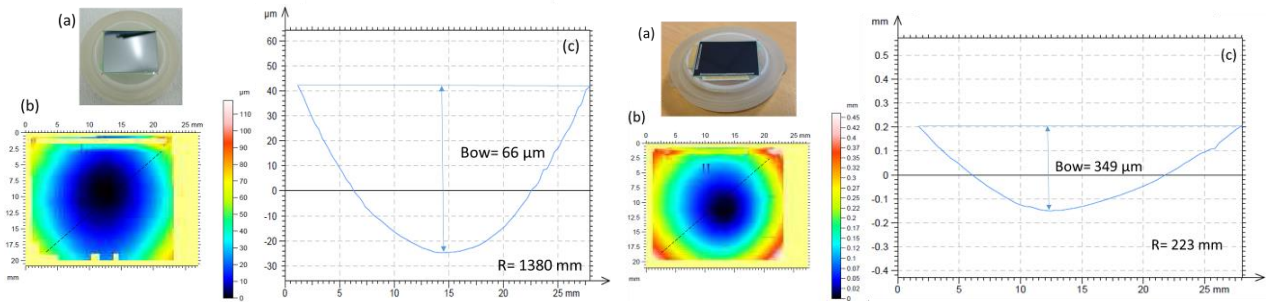

Figure 11. Left: (a) Silicon die sample on flexible packaging in flat position; (b) map characterization; (c) altitude profile and radius of curvature extraction in flat position according to the dotted line in (b).
Right: Silicon die sample on flexible packaging in curved position; (b) map characterization; (c) altitude profile and radius of curvature extraction in curved position according to the dotted line in (b).

For CMOS sensor with PCB and electrical connections, the same method was applied and a 280 mm radius of curvature was characterized. Flat reference and curved position are respectively illustrated in figure 11.a and 11.b. Concave curved position shape is characterized in figure 11.c.

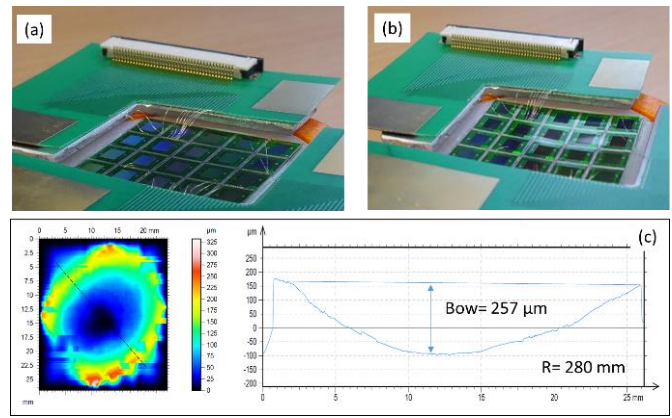

Figure 12. (a) 20*23*0.1 mm CMOS image sensor packaged on T-CFPA packaging in flat position; (b) the same sample in concave curved position; (c) altitude profile and radius of curvature extraction in curved position according to the dotted line.

## 4. CONCLUSION

A tunable curvature packaging for large CMOS image sensor in spherical concave and convex shape has been proposed, optimized, and tested on 20*23*0.1 mm CMOS sensors samples.

Based on optical advantages of a curved focal plane array on the optical design simplification, development of simulation tools and analytical models led to understand and determine the spherical bending limits of thinned CMOS image sensors. Buckling effect at the middle of the edges, due to the spherical shape, could be one particular bending limitation. An estimation of the maximum stress level allows to estimate the maximum achievable curvature. For a 30*30*0.1mm Silicon substrate a radius of curvature higher than 130mm is achievable. A dedicated Tunable Curved Focal Plane Area (T-CFPA) packaging including optimized mechanical interface and electrical wire bonding has been realized on 3D printing tool with a 20*23*0.1 mm CMOS image sensor. Topology characterizations highlight the concave bending capability and reversibility of the T-CFPA packaging down to R=280 mm.

Further development will include the bending influence on electro-optical performances. Also, a fisheye prototype realization with curved sensor is in progress to validate this new optical design approach.

## ACKNOWLEDGEMENTS

This activity was partially funded by the French Research Agency (ANR) under the program ANR-OASIX #12-JS05-0004 as well as the LabEx FOCUS ANR-11-LABX-0013, and will be pursued in the frame of the H2020 - ERC-STG-2015 – 678777 ICARUS program over the period 2016-2021.